\documentclass[9pt]{article}
\usepackage{spconf,amsmath,graphicx,amssymb,amsthm}
\usepackage{algorithm, algorithmic}
\usepackage{subfig}

\newcommand{\R}{\mathbb{R}}

\def\minim{\mathop{\hbox{minimize}}}

\theoremstyle{remark}

\theoremstyle{definition}

\title{BEYOND $\ell_1$-norm MINIMIZATION FOR SPARSE SIGNAL RECOVERY}
\name{Hassan Mansour \thanks{hassanm@cs.ubc.ca}\vspace{-0.2in}
\thanks{The author was supported in part by the Natural
Sciences and Engineering Research Council of Canada (NSERC)
Collaborative Research and Development Grant DNOISE II
(375142-08).}}
\address{University of British Columbia, Vancouver - BC, Canada}
\begin{document}
%\ninept
%
\maketitle
\begin{abstract}
Sparse signal recovery has been dominated by the basis pursuit denoise (BPDN) problem formulation for over a decade. In this paper, we propose an algorithm that outperforms BPDN in finding sparse solutions to underdetermined linear systems of equations at no additional computational cost. Our algorithm, called WSPGL1, is a modification of the spectral projected gradient for $\ell_1$ minimization (SPGL1) algorithm in which the sequence of LASSO subproblems are replaced by a sequence of weighted LASSO subproblems with constant weights applied to a support estimate. The support estimate is derived from the data and is updated at every iteration. The algorithm also modifies the Pareto curve at every iteration to reflect the new weighted $\ell_1$ minimization problem that is being solved. We demonstrate through extensive simulations that the sparse recovery performance of our algorithm is superior to that of $\ell_1$ minimization and approaches the recovery performance of iterative re-weighted $\ell_1$ (IRWL1) minimization of Cand{\`e}s, Wakin, and Boyd, although it does not match it in general.  Moreover, our algorithm has the computational cost of a single BPDN problem. 
\end{abstract}
%

%%%%%%%%%%%
\section{Introduction}
\label{sec:intro}

The problem of recovering a sparse signal from an underdetermined system of linear equations is prevalent in many engineering applications. In fact, this problem has given rise to the field of compressed sensing which presents a new paradigm for acquiring signals that admit sparse or nearly sparse representations using fewer linear measurements than their ambient dimension \cite{Donoho2006_CS, CRT05}. 

Consider an arbitrary signal $x \in \R^N$ and let $y \in \R^n$ be a
set of measurements given by $y = Ax + e,$ where $A$ is a known
$n\times N$ measurement matrix, and $e$ denotes additive noise that
satisfies $\|e\|_2\leq \epsilon$ for some known $\epsilon\ge
0$. Compressed sensing theory states that it is possible to recover
$x$ from $y$ (given $A$) even when $n \ll N$, that is, using very few
measurements. When $x$ is strictly sparse---i.e., when there are only
$k < n$ nonzero entries in $x$---and when $e=0$, one may recover an
estimate $\hat{x}$ of the signal $x$ by solving the constrained $\ell_0$
minimization problem
\begin{equation}\label{eq:L0_min}
\minim_{u\in \R^N}\ \|u\|_0 \ \text{subject to} \ \ Au=y.
\end{equation}
However, $\ell_0$ minimization is a combinatorial
problem and quickly becomes intractable as the dimensions
increase. Instead, the convex relaxation given by the $\ell_1$ minimization problem
$$\minim_{u \in \R^N}\ \|u\|_1 \ \text{subject to} \ \|Au - y\|_2 \leq \epsilon \qquad \textrm{(BPDN)}$$ also known as \emph{basis pursuit denoise} (BPDN) \cite{chen99atomic}, can be used to
recover an estimate $\hat{x}$.  Cand{\'e}s, Romberg and Tao \cite{CRT05}
and Donoho \cite{Donoho2006_CS} show that it is possible to recover a stable and robust approximation of $x$ by solving (BPDN) instead of \eqref{eq:L0_min} at the cost of increasing
the number of measurements taken.

%
%Cand{\'e}s, Romberg and Tao \cite{CRT05}
%and Donoho \cite{Donoho2006_CS} show that (BP) can stably and robustly recover $x$ from inaccurate and what
%appears to be ``incomplete'' measurements $y = Ax + e$ if $A$ is an
%appropriate measurement matrix, e.g., a Gaussian random matrix such
%that $n \gtrsim k\log(N/k)$. Contrary to $\ell_0$ minimization, (BPDN)
%is a convex program and can be solved efficiently. Consequently, it is
%possible to recover a stable and robust approximation of $x$ by
%solving (BPDN) instead of \eqref{eq:L0_min} at the cost of increasing
%the number of measurements taken.

Several works in the literature have proposed alternate algorithms
that attempt to bridge the gap between $\ell_0$ and $\ell_1$
minimization. These include using $\ell_p$ minimization with $0 < p <
1$ which has been shown to be stable and robust under weaker
conditions than those of $\ell_1$ minimization, see
\cite{gribonval07:_highl, chartrand2008rip, Saab_ellp:2010}.
\emph{Weighted $\ell_1$ minimization} is another alternative
if there is prior information regarding the support of the signal
to-be-recovered as it incorporates such information into the recovery
by weighted basis pursuit denoise (w-BPDN) 
$$\minim_{u}\ \|u\|_{1,\mathrm{w}}\ \text{subject to}\ \|Au - y\|_2 \leq \epsilon, \quad \textrm{(w-BPDN)}$$
where $\mathrm{w}\in (0,1]^N$ and $\|u\|_{1,\mathrm{w}} := \sum_i
\mathrm{w}_i |u_i|$ is the weighted $\ell_1$ norm (see
\cite{CS_using_PI_Borries:2007,
  Vaswani_Lu_Modified-CS:2010,FMSY:2011}). 
  
 When no prior information is available, the \emph{iterative reweighted $\ell_1$ minimization} (IRWL1) algorithm,
proposed by Cand\`{e}s, Wakin, and Boyd \cite{candes2008irl1} and
studied by Needell \cite{Needell:2009}, solves a sequence of weighted
$\ell_1$ minimization problems with the weights $\mathrm{w}_i^{(t)}
\approx 1/\left|x_i^{(t-1)}\right|$, where $x_i^{(t-1)}$ is the
solution of the $(t-1)$th iteration and $\mathrm{w}_i^{(0)} = 1$ for
all $i \in \{1\dots N\}$. More recently, Mansour and Yilmaz \cite{Mansour:ICASSP12a} proposed a \textit{support driven iterative reweighted $\ell_1$ minimization} (SDRL1) algorithm that also solves a sequence of weighted $\ell_1$ minimization problems with constant weights $\mathrm{w}_i^{(t)} = \omega \in [0,1]$ when $i$ belongs to support estimates $\Lambda^{(t)}$ that are updated in every iteration. The performance of SDRL1 is shown to match that of IRWL1.

Motivated by the performance of constant weighting in the SDRL1 algorithm, we present in this paper an iterative algorithm called WSPGL1 that converges to the solution of a weighted $\ell_1$ problem (wBPDN) with a two set weight vector $\mathrm{w}_{\Lambda} = \omega$ and $\mathrm{w}_{\Lambda^c} = 1$, where $\omega \in [0,1]$ and $\Lambda$ is a support estimate. The set $\Lambda$ to which the algorithm converges is not known a priori but is derived and updated at every iteration. Our algorithm is a modification of the \textit{spectral projected gradient for $\ell_1$ minimization} (SPGL1) algorithm \cite{BergFriedlander:2008} which solves a sequence of LASSO \cite{tibshirani:96} subproblems to arrive at the solution of the BPDN problem. We give an overview of the SPGL1 algorithm in section \ref{sec:SPGL1}. In contrast, our algorithm solves a sequence of weighted LASSO subproblems that converge to the solution of the wBPDN problem with weights $\omega$ applied to a support estimate $\Lambda$. We discuss the details of this algorithm in section \ref{sec:WSPGL1} and present preliminary recovery results in section \ref{sec:Results} demonstrating its superior performance in recovering sparse signals from incomplete measurements compared with $\ell_1$ minimization. We limit the scope of this paper to discussing the algorithm and presenting sparse recovery results and leave the analysis of the algorithm for future work.\\

\textbf{Notation:} For a vector $x \in \R^N$, an index set $\Lambda \subset \{1\dots N\}$ and its complement $\Lambda^c$, let $x_k$ and $x|_k$ refer to the largest $k$ entries of $x$, $x(k)$ is the $k$th largest entry of $x$, $x_{\Lambda}$ refers to the entries of $x$ indexed by $\Lambda$, and $x^{(t)}$ is the vector $x$ at iteration $t$.

%%%%%%%%%%%%%%%%%%%%
\section{The SPGL1 algorithm}\label{sec:SPGL1}
In this section, we give an overview of the SPGL1 algorithm, developed by
van den Berg and Friedlander \cite{BergFriedlander:2008}, that finds the solution to the BPDN problem.

\subsection{General overview}
The SPGL1 algorithm finds the solution of the BPDN problem by efficiently solving a sequence of LASSO subproblems
$$\minim_{u \in \R^N}\ \|Au - y\|_2 \ \text{subject to} \  \|u\|_1 \leq \tau \qquad (\textrm{LS}_{\tau})$$
using a spectral projected-gradient algorithm. The single parameter $\tau$ determines a Pareto curve $\phi(\tau) = \|r^{\tau}\|_2$, where $r^{\tau} = y - Ax^{\tau}$ and $x^{\tau}$ is the solution of (LS$_{\tau}$). The Pareto curve traces the optimal trade-off between the least-squares fit and the one-norm of the solution.

The SPGL1 algorithm is initialized at a point $x^{(0)}$ which gives an initial $\tau_0 = \|x^{(0)}\|_1$. The parameter $\tau$ is then updated according to the following rule
\begin{equation}\label{eq:tau_update}
	\tau_{t+1} = \tau_{t} + \frac{\|r^{\tau_t}\|_2 - \epsilon}{\|A^Hr^{\tau_t}\|_{\infty}/\|r^{\tau_t}\|_2},
\end{equation}
where superscript $^H$ indicates Hermitian transpose, and $\epsilon = \|e\|_2 = \|y-Ax\|_2$. Consequently, the next iterate $x^{(t+1)}$ is given by the solution of (LS$_{\tau_{t+1}}$) and the algorithm proceeds until convergence.

\subsection{Probing the Pareto curve}
One of the main contributions of \cite{BergFriedlander:2008} lies in recognizing and proving that the Pareto curve is convex and continuously differentiable over all solutions of (LS$_{\tau}$). This gives rise to the update rule for $\tau$ shown in \eqref{eq:tau_update} and guarantees the convergence of SPGL1 to the solution of BPDN. 

\begin{figure}
	\includegraphics[width=3in]{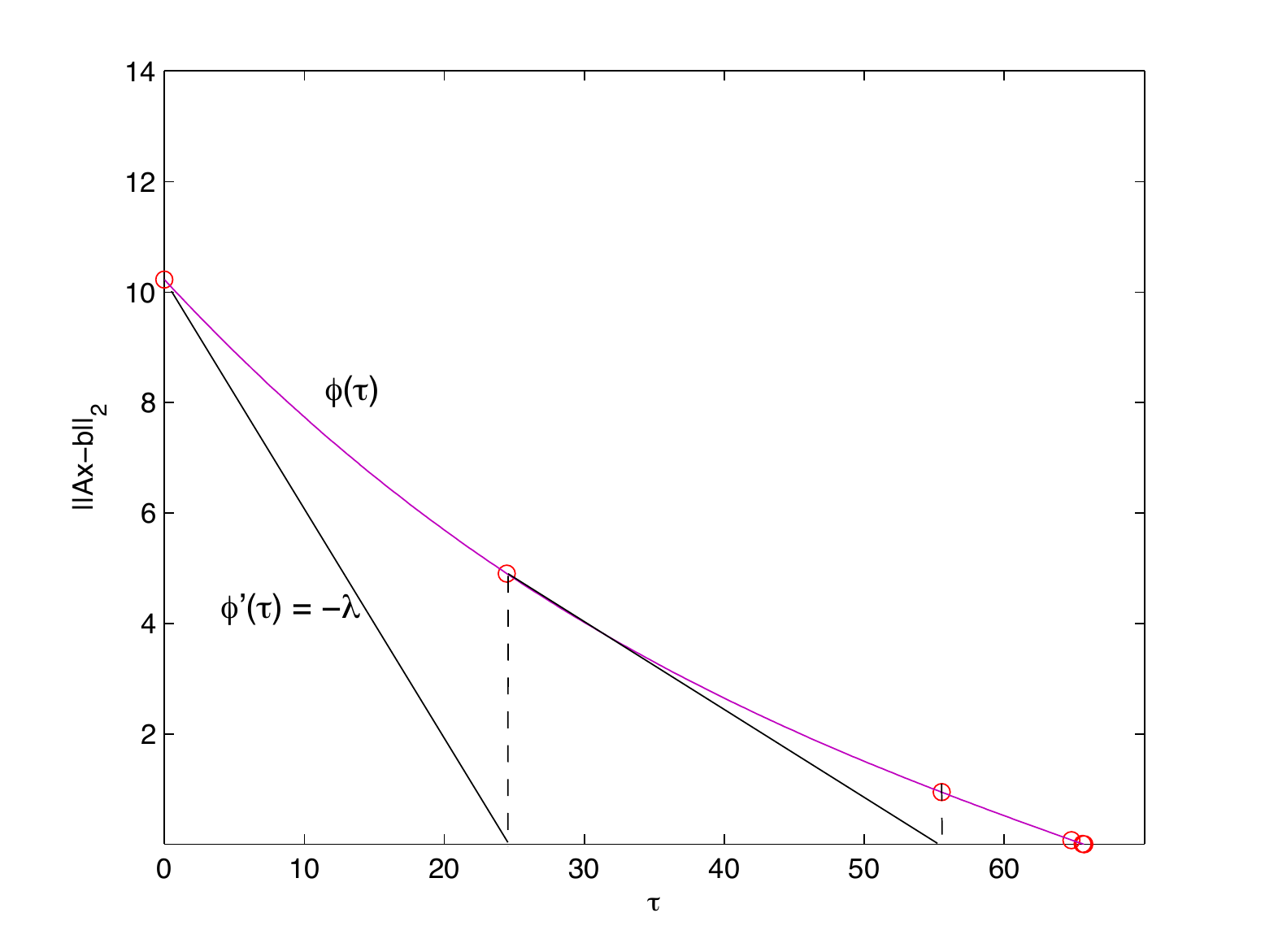}
	\caption{Example of a typical Pareto curve showing the root finding iterations used in SPGL1 \cite{BergFriedlander:2008}.}
	\label{fig:pareto}
\end{figure}

The update rule \eqref{eq:tau_update} is in fact a Newton-based root-finding method that solves $\phi(\tau) = \epsilon$. The update rule generates a sequence of parameters $\tau_t$ according to the Newton iteration
$$
	\tau_{t+1} = \tau_{t} + \frac{\epsilon - \phi(\tau_t)}{\phi'(\tau_t)}
$$
where $\phi'(\tau_t)$ is the derivative of $\phi$ at $\tau_t$. It is then shown that the $\phi'(\tau)$ is equal to the negative of the dual variable $\lambda$ of (LS$_{\tau}$) resulting in the expression $\phi'(\tau) = -\lambda = -\frac{\|A^Hr\|_{\infty}}{\|r\|_2}$. Figure \ref{fig:pareto} illustrates an example of a Pareto curve and the root finding method used in SPGL1.

%%%%%%%%%%%
\section{The proposed WSPGL1 algorithm}\label{sec:WSPGL1}

In this section, we describe the proposed WSPGL1 algorithm for sparse signal recovery as a variation of the SPGL1 algorithm. The WSPGL1 algorithm solves a sequence of weighted LASSO subproblems to arrive at the solution to a weighted BPDN problem with weights $\omega \in [0,1]$ applied to a support set $\Lambda$. The set $\Lambda$ is derived and updated from the solutions of the weighted LASSO subproblems (LS$_{\tau}, \mathrm{w}$).

\subsection{Algorithm description}
The two algorithms SPGL1 and WSPGL1 follow exactly the same initial steps until the solution $x^{\tau_1}$ of the first LASSO problem (LS$_{\tau_1}$) is found. At this point, WSPGL1 generates a support set $\Lambda$ containing the support of the $k$ largest in magnitude entries of $x^{\tau_1}$. A weight vector $\mathrm{w}$ is then generated such that
$$\mathrm{w}_i = \left\{
\begin{array}{l}
	\omega, \quad i \in \Lambda \\
	1, \quad i \in \Lambda^c 
\end{array} \right.
$$
We heuristically choose $k = n/\left(2\log(N/n)\right)$ and $\omega = 0.3$. 

The weight vector is then used to define the weighted LASSO subproblem
$$\minim_{u \in \R^N}\ \|Au - y\|_2 \ \text{subject to} \  \|u\|_{1,\mathrm{w}} \leq \tau \qquad (\textrm{LS}_{\tau,\mathrm{w}})$$
with the corresponding dual variable $$\lambda_{\mathrm{w}} = \frac{\|A^Hr\|_{\infty,\mathrm{w}}}{\|r\|_2},$$ 
where $\|v\|_{\infty,\mathrm{w}} = \|v \cdot \mathrm{w}^{-1}\|_{\infty}$. The weighted LASSO subproblem and its dual constitute a subproblem of (wBPDN) with support estimate $\Lambda$. The BPDN and wBPDN problems have different Pareto curves. Therefore, the iterate $(\|r^{1}\|_2,\tau_1)$ which lies on the Pareto curve of BPDN must be adjusted to lie on the Pareto curve of the wBPDN problem. This can be easily achieved by switching $\tau_1$ with $\tau_1' = \|x^{\tau_1}\|_{1,\mathrm{w}}$. The WSPGL1 algorithm then proceeds according to the following pseudocode.

\begin{algorithm}\caption{The WSPGL1 algorithm}
\begin{algorithmic}[1]\label{alg:WSPGL1}
\STATE \textbf{Input} $y = Ax + e$, $\epsilon$, $k = n/\left(2\log(N/n)\right)$, $\omega \in [0,1]$
\STATE \textbf{Output} $x^{(t)}$
\STATE \textbf{Initialize} $\mathrm{w}_i^{(0)} = 1$ for all $i \in \{1\dots N\}$\\
\quad\quad\quad\quad $t = 0$, $x^{(0)} = 0$, $\tau_0 = 0$  
\LOOP
\STATE $t = t + 1$
\STATE $\Lambda = \textrm{supp}(x^{(t-1)}|_k)$, ${\small
\mathrm{w}_i = \left\{
\begin{array}{l}
	\omega, \quad i \in \Lambda \\
	1, \quad i \in \Lambda^c \\
\end{array} \right.}
$
\STATE $\tau_{t-1}' = \|x^{(t-1)}\|_{1,\mathrm{w}}$ \vspace{0.05in}
\STATE  {\large$ \tau_{t} = \tau_{t-1}'+ \frac{\|r^{\tau_{t-1}}\|_2 \quad-\quad  \epsilon}{\|A^Hr^{\tau_{t-1}}\|_{\infty,\mathrm{w}}/\|r^{\tau_{t-1}}\|_2} $}\vspace{0.05in}
\STATE $x^{(t)} = \arg\min\limits_{u} \|Au - y\|_2$  s.t. $ \|u\|_{1,\mathrm{w}} \leq \tau_{t}$
\STATE $r^{\tau_{t}} = y - Ax^{(t)}$
\ENDLOOP
\end{algorithmic}
\end{algorithm}\vspace{-0.2in}

\subsection{Discussion}
\begin{figure}[t]
	\mbox{
	\subfloat[]{\includegraphics[width=3in, height=2.2in]{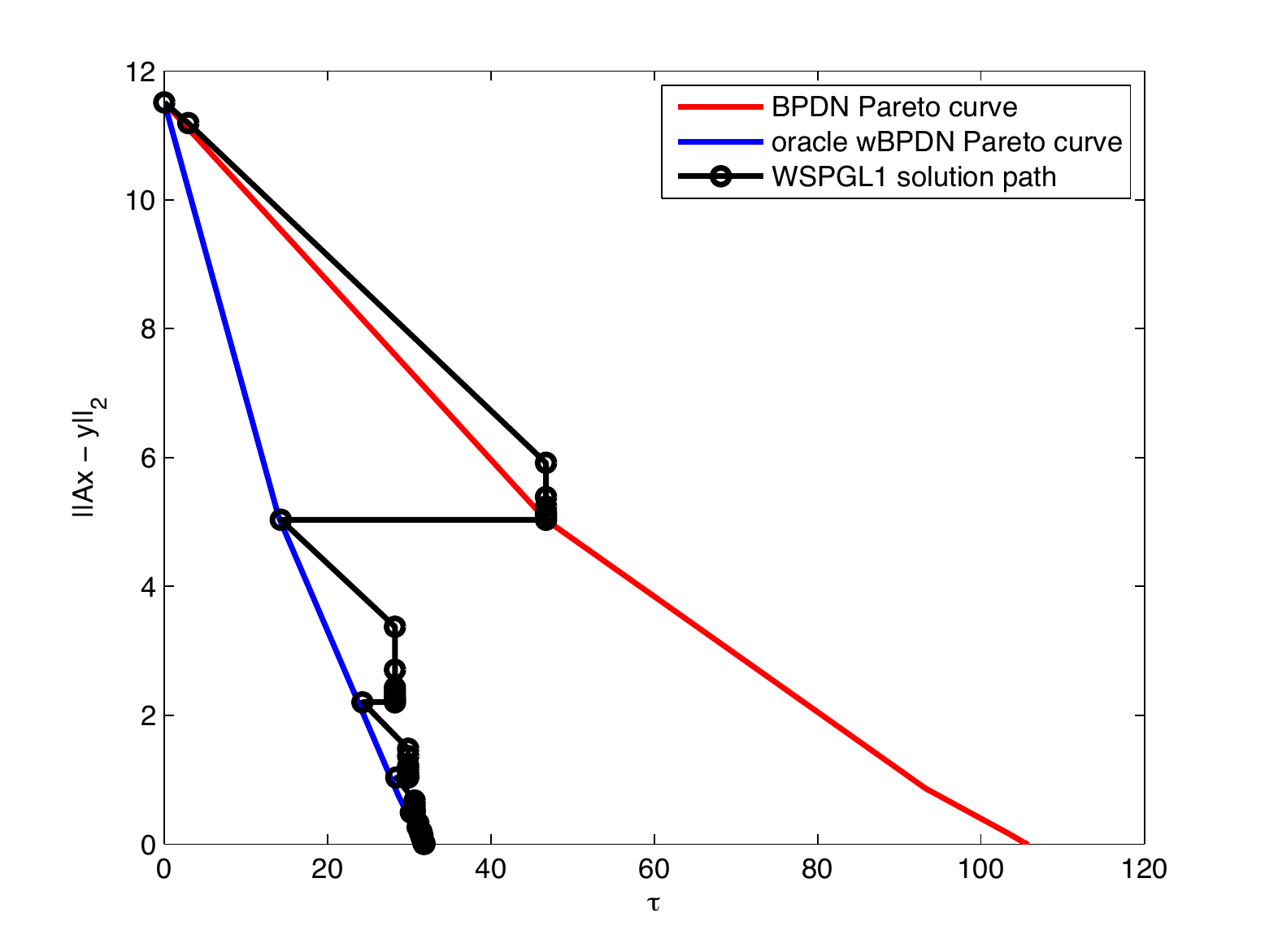}}}
	\mbox{
	\subfloat[]{\includegraphics[width=3in, height=2.2in]{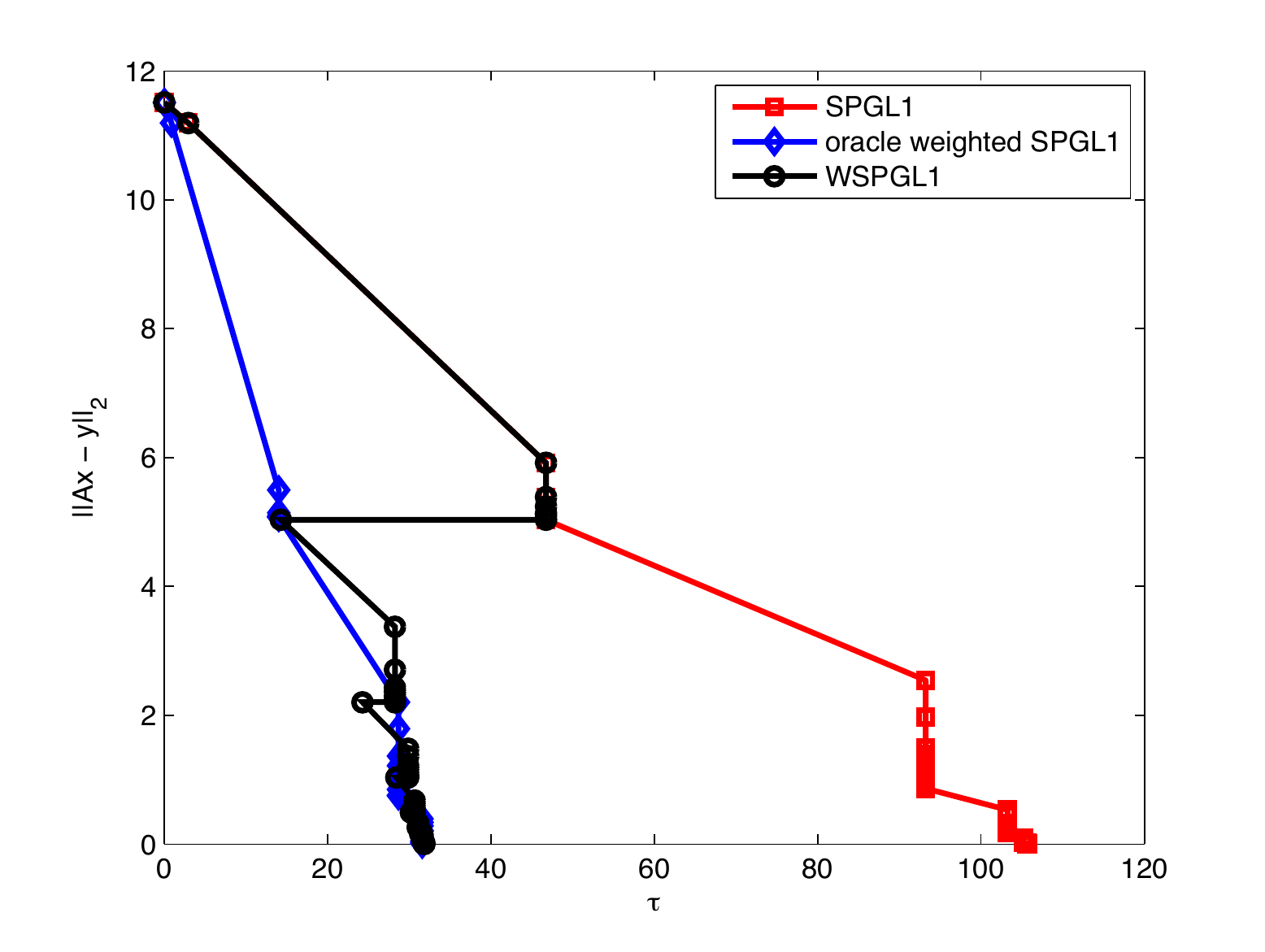}}}
	\caption{(a) The solution path for WSPGL1 follows the BPDN Pareto curve until the first (LS$_{\tau}$) is solved, after which WSPGL1 switches to the wBPDN Pareto curve. (b) Solution paths of WSPGL1, SPGL1, and weighted SPGL1 with oracle support information. Both WSPGL1 and the oracle weighted SPGL1 use $\omega = 0.3$. } \vspace{-0.2in}
	\label{fig:solution_path}
\end{figure}

The WSPGL1 algorithm converges to the solution of a weighted BPDN problem with weights $\omega \in [0,1]$ applied to a support set $\Lambda$. When the sparse signal is recovered exactly, the set $\Lambda$ coincides with the true support of the sparse signal $x$. Figure \ref{fig:solution_path} (a) illustrates the solution path of WSPGL1 which follows the Pareto curve of the BPDN problem until the first (LS$_{\tau}$) is solved. The algorithm then uses the support information from $x^{\tau_1}$ to switch to the Pareto curve of the wBPDN problem. Figure \ref{fig:solution_path} (b) compares the solution paths of WSPGL1, SPGL1, and oracle weighted SPGL1 with weight $\omega = 0.3$ applied to the true signal support. It can be seen that WSPGL1 converges to the solution of the oracle weighted $\ell_1$ problem. Moreover, the solution paths of these algorithms merge after only the first (LS$_{\tau}$) subproblem. Note here that the x-axis is the parameter $\tau$ which is equal to the one-norm of $x^{(t)}$ for SPGL1 and the weighted one-norm of $x^{(t)}$ for WSPGL1 and the oracle weighted SPGL1.

\begin{figure*}[ht]
	\centering
	\includegraphics[width = 7in, height=2in]{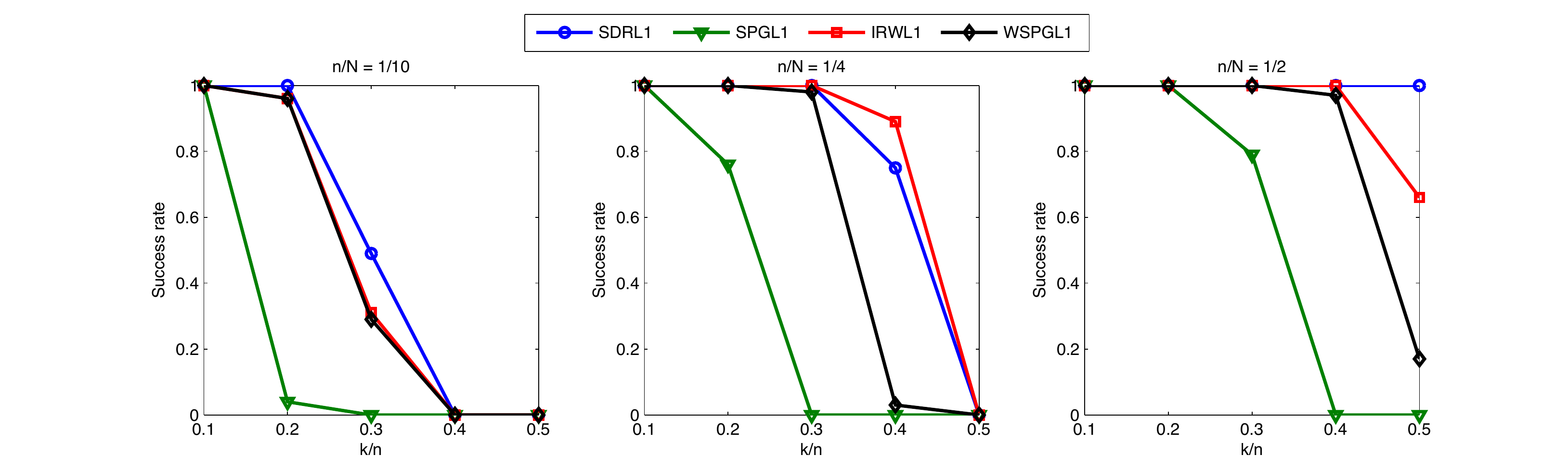}
	\caption{Comparison of the percentage of exact recovery of \emph{sparse signals} between the proposed WSPGL1, SDRL1 \cite{Mansour:ICASSP12a}, IRL1 \cite{candes2008irl1}, and standard $\ell_1$ minimization using SPGL1 \cite{BergFriedlander:2008}. The signals have an ambient dimension $N = 2000$ and the sparsity and number of measurements are varied. The results are averaged over 100 experiments.}\label{fig:sparse_recovery}
\end{figure*}

It is still not clear under what conditions the WSPGL1 algorithm achieves exact recovery. What is clear is that WSPGL1 can exactly recover signals with far more nonzero coefficients than what BPDN can recover.  The WSPGL1 algorithm is motivated by the work in \cite{FMSY:2011} and \cite{Mansour:ICASSP12a}, which show that weighted $\ell_1$ minimization can recover less sparse signals than BPDN when the weights are applied to a support estimate that is at least 50\% accurate. Moreover, it is possible to draw a support estimate from the solution of BPDN and improve that support estimate by solving wBPDN using the initial support estimate. Based on these results, we conjectured that the solution of every LASSO subproblem in SPGL1 allows us to find a support estimate that is accurate enough to improve the recovery conditions of the corresponding wBPDN problem.  A full analysis of this algorithm will be the subject of future work. \vspace{-0.1in}

%
%\begin{figure*}[t]
%	\centering
%	\includegraphics[width = 6in, height=1.5in]{iter_wL1_comparison_compr_hist_n-N_0-1.pdf}
%	\caption{Histogram of the ratio of the mean squared error (MSE) between the proposed SDRL1 and IRL1 \cite{candes2008irl1} for the recovery of \emph{compressible signals}. The signals $x$ follow a power law decay such that $|x_i| = ci^{-p}$, for constant $c$ and exponent $p$.}\label{fig:compressible_histogram}
%\end{figure*}

%%%%%%%%%%%%%%%%%%%%%%%%
\section{Numerical results}\label{sec:Results}
We tested the WSPGL1 algorithm by comparing its performance with SDRL1 \cite{Mansour:ICASSP12a}, IRWL1 \cite{candes2008irl1} and standard $\ell_1$ minimization using the SPGL1 \cite{BergFriedlander:2008} algorithm in recovering synthetic signals $x$ of dimension $N = 2000$. We first recover sparse signals from compressed measurements $y = Ax$ using matrices $A$ with i.i.d. Gaussian random entries and dimensions $n
\times N$ where $n \in \{N/10, N/4, N/2 \}$. The sparsity of the signal is varied such that $k/n \in \{0.1, 0.2, 0.3, 0.4, 0.5\}$. To quantify the reconstruction performance, we plot in Figure \ref{fig:sparse_recovery} the percentage of successful recovery averaged over 100
realizations of the same experimental conditions. The figure shows that in all cases, the WSPGL1 algorithm outperforms standard $\ell_1$ minimization in recovering sparse signals. Moreover, the recovery performance approaches that of the iterative reweighted $\ell_1$ algorithms SDRL1 and IRWL1 while requiring only a fraction of the computational cost associated with these algorithms. 
\vspace{-0.2in}

%Next, we generate compressible signals with power law decay such that $x(i) = ci^{-p}$ for some constant $c$ and decay power $p$. We consider the case where $n/N = 0.1$ and the decay power $p \in \{1.1, 1.5, 2\}$ and plot the ratio of the reconstruction error of SDRL1 over that of IRL1. Figure \ref{fig:compressible_histogram} shows the histograms of the ratio for 100 experiments each. Note that a ratio smaller than one means that our algorithm has a smaller reconstruction error than that of IRL1. The histograms indicate that both algorithms have a comparable performance for signals with different decay rates.

\small
\bibliographystyle{IEEEbib}
\bibliography{sparse}

\begin{thebibliography}{10}

\bibitem{Donoho2006_CS}
D.~Donoho,
\newblock ``Compressed sensing.,''
\newblock {\em IEEE Transactions on Information Theory}, vol. 52, no. 4, pp.
  1289--1306, 2006.

\bibitem{CRT05}
E.~J. Cand{\`e}s, J.~Romberg, and T.~Tao,
\newblock ``Stable signal recovery from incomplete and inaccurate
  measurements,''
\newblock {\em Communications on Pure and Applied Mathematics}, vol. 59, pp.
  1207--1223, 2006.

\bibitem{chen99atomic}
S.~Chen, D.~Donoho, and M.A. Saunders,
\newblock ``Atomic decomposition by basis pursuit,''
\newblock {\em SIAM Journal on Scientific Computing}, vol. 20, no. 1, pp.
  33--61, 1999.

\bibitem{gribonval07:_highl}
R.~Gribonval and M.~Nielsen,
\newblock ``Highly sparse representations from dictionaries are unique and
  independent of the sparseness measure,''
\newblock {\em Applied and Computational Harmonic Analysis}, vol. 22, no. 3,
  pp. 335--355, May 2007.

\bibitem{chartrand2008rip}
R.~Chartrand and V.~Staneva,
\newblock ``{Restricted isometry properties and nonconvex compressive
  sensing},''
\newblock {\em Inverse Problems}, vol. 24, no. 035020, 2008.

\bibitem{Saab_ellp:2010}
R.~Saab and O.~Yilmaz,
\newblock ``Sparse recovery by non-convex optimization -- instance
  optimality,''
\newblock {\em Applied and Computational Harmonic Analysis}, vol. 29, no. 1,
  pp. 30--48, July 2010.

\bibitem{CS_using_PI_Borries:2007}
R.~von Borries, C.J. Miosso, and C.~Potes,
\newblock ``Compressed sensing using prior information,''
\newblock in {\em 2nd IEEE International Workshop on Computational Advances in
  Multi-Sensor Adaptive Processing, CAMPSAP 2007.}, 12-14 2007, pp. 121 -- 124.

\bibitem{Vaswani_Lu_Modified-CS:2010}
N.~Vaswani and Wei Lu,
\newblock ``{Modified-CS}: Modifying compressive sensing for problems with
  partially known support,''
\newblock {\em arXiv:0903.5066v4}, 2009.

\bibitem{FMSY:2011}
M.~P. Friedlander, H.~Mansour, R.~Saab, and \"{O}. Y{\i}lmaz,
\newblock ``Recovering compressively sampled signals using partial support
  information,''
\newblock {\em to appear in the IEEE Trans. on Inf. Theory}.

\bibitem{candes2008irl1}
E.~J. Cand{\`e}s, Michael~B. Wakin, and Stephen~P. Boyd,
\newblock ``Enhancing sparsity by reweighted $\ell_1$ minimization,''
\newblock {\em The Journal of Fourier Analysis and Applications}, vol. 14, no.
  5, pp. 877--905, 2008.

\bibitem{Needell:2009}
D.~Needell,
\newblock ``Noisy signal recovery via iterative reweighted l1-minimization,''
\newblock in {\em Proceedings of the 43rd Asilomar conference on Signals,
  systems and computers}, 2009, Asilomar'09, pp. 113--117.

\bibitem{Mansour:ICASSP12a}
{H. Mansour} and O.~Yilmaz,
\newblock ``Support driven reweighted $\ell_1$ minimization,''
\newblock in {\em Proc. of the IEEE International Conference on Acoustics,
  Speech, and Signal Processing (ICASSP)}, March 2012.

\bibitem{BergFriedlander:2008}
E.~van~den Berg and M.~P. Friedlander,
\newblock ``Probing the pareto frontier for basis pursuit solutions,''
\newblock {\em SIAM Journal on Scientific Computing}, vol. 31, no. 2, pp.
  890--912, 2008.

\bibitem{tibshirani:96}
R.~Tibshirani,
\newblock ``{Regression shrinkage and selection via the lasso},''
\newblock {\em J. Roy. Statist. Soc. Ser. B}, vol. 58, no. 1, pp. 267--288,
  1996.

\end{thebibliography}

\end{document}